\documentclass[aps,twocolumn,showpacs,superscriptaddress,floatfix]{revtex4}
\usepackage[T1]{fontenc}
\usepackage{graphicx}
\usepackage{ae}

\newcommand{\kB}   {k_B}
\newcommand{\ldb}  {\lambda_{\rm dB}}

\begin{document}

\title{Experimental observations of density fluctuations in an elongated Bose gas:\\
ideal gas and quasi-condensate regimes}

\author{J.~Esteve}
\email{esteve@matterwave.de}
\homepage{http://atomoptic.iota.u-psud.fr/}
\affiliation{Laboratoire Charles Fabry, CNRS et Université Paris Sud 11, 91403 Orsay CEDEX, France}
\affiliation{Laboratoire de Photonique et de Nanostructures, CNRS, 91460 Marcoussis, France}
\author{J.-B.~Trebbia}
\affiliation{Laboratoire Charles Fabry, CNRS et Université Paris Sud 11, 91403 Orsay CEDEX, France}
\author{T.~Schumm}
\affiliation{Laboratoire Charles Fabry, CNRS et Université Paris Sud 11, 91403 Orsay CEDEX, France}
\author{A.~Aspect}
\affiliation{Laboratoire Charles Fabry, CNRS et Université Paris Sud 11, 91403 Orsay CEDEX, France}
\author{C. I.~Westbrook}
\affiliation{Laboratoire Charles Fabry, CNRS et Université Paris Sud 11, 91403 Orsay CEDEX, France}
\author{I.~Bouchoule}
\affiliation{Laboratoire Charles Fabry, CNRS et Université Paris Sud 11, 91403 Orsay CEDEX, France}

\begin{abstract}
We report \emph{in situ} measurements of density fluctuations in a
quasi one dimensional $^{87}$Rb Bose gas at thermal equilibrium in
an elongated harmonic trap. We observe an excess of fluctuations
compared to the shot noise level expected for uncorrelated atoms. At
low atomic density, the measured excess is in good agreement with
the expected ``bunching'' for an ideal Bose gas. At high density, the
measured fluctuations are strongly reduced compared to the ideal gas
case. We attribute this reduction to repulsive inter-atomic
interactions. The data are compared with a calculation for an
interacting Bose gas in the quasi-condensate regime.
\end{abstract}

\maketitle

In a classical gas, the mean square fluctuation of the number of particles within a small volume is equal to the number of particles (we shall call this fluctuation ``shot noise''). On the other hand, because of quantum effects, the fluctuations in a non-condensed Bose gas are larger than the shot noise contribution~\cite{Landau}. For photons, the well known Hanbury Brown-Twiss or ``photon bunching'' effect is an illustration of this phenomenon~\cite{BunchingOptique}. Analogous studies have been undertaken to measure correlations between bosonic atoms released from a trap after a time of flight~\cite{Yasuda96, Helium, Folling2005, Esslinger}. However, bunching in the density distribution of trapped cold atoms at thermal equilibrium has not been yet directly observed. 

Density fluctuations of a cold atomic sample can be measured by absorption imaging as proposed in~\cite{Grondalski99,Altman2004} and recently shown in~\cite{Folling2005, Greiner2005}. When using this method, one necessarily integrates the density distribution over one direction, and this integration can mask the bunching effect whose correlation length is of the order of the de~Broglie wavelength. A one dimensional (1D) gas, {\emph i.e.} a gas in an anisotropic confining potential with a temperature lower than or of order of the zero point energy in two directions, allows one to avoid this integration, and is thus a very favorable experimental geometry. 

Additionally, atoms in 1D do not Bose condense~\cite{Hohenberg:1967}. One can therefore achieve a high degree of quantum degeneracy without condensation, which enhances the bunching effect for an ideal gas. When one considers the effect of interactions between atoms, two additional regimes can appear: the Tonks-Girardeau regime and the quasi-condensate regime~\cite{kheruntsyan:053615}. Starting from an ideal gas, as one increases density at fixed temperature $T$, the 1D interacting Bose gas passes smoothly to the quasi-condensate regime. The linear density scale for this crossover is given by $n_{T} = (m (\kB T)^2/\hbar^2 g)^{1/3}$ where $g$ is the effective 1D coupling constant and $m$ the atomic mass~\cite{Kheru2003,Quasibec_Castin}. Density fluctuations are suppressed by a factor $(n/n_T)^{3/2}$ compared to the ideal gas (see Eq.~(4) below), although phase fluctuations remain~\cite{Dettmer2002, richard:010405, Helleg2003, Hugbart2005, shvarchuck2002}.  We emphasize that this crossover occurs in the dense, weakly interacting limit which is the opposite of the Tonks-Girardeau regime.

To measure the density fluctuations of a trapped Bose gas as a function of its density, we acquire a large number of images of different trapped samples under identical conditions.  We have access to both the ideal Bose gas limit, in which we observe the expected excess fluctuations compared to shot-noise, as well as the quasi-condensate regime in which repulsive interactions suppress the density fluctuations. 

\begin{figure}
\includegraphics{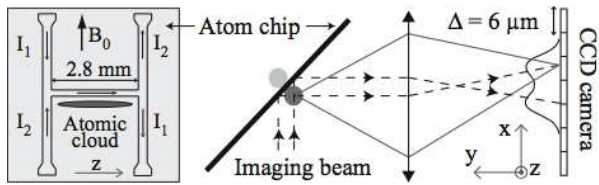}
\caption{Schematic of the experimental setup. Left: drawing of the
wires constituting the atom chip. We keep $I_1+I_2=3$~A and adjust
$I_1-I_2$ between 0.3~A and 1~A to vary the confinement along $z$.
The uniform field $B_0$ is approximately 40~G, we also add a small
field ($\lesssim 1$~G) along $z$. Right: optical imaging system. We
image the cloud and its reflection on the atom chip onto a CCD
camera. In the radial direction, the unresolved cloud images cover
approximately five pixels whose size in the object plane is
$\Delta=6$~$\mu$m. } \label{fig.dispositif}
\end{figure}

Our measurements are conducted in a highly anisotropic magnetic trap
created by an atom chip. We use three current carrying wires forming
an H pattern~\cite{reichel:2002} and an external uniform magnetic field to magnetically trap the $^{87}$Rb atoms in the $|F=2,m_F=2\rangle$ state (see Fig.~\ref{fig.dispositif}). Adjusting the currents in the wires and the external magnetic field, we can tune the longitudinal frequency between 7 and 20~Hz while keeping the transverse frequency $\omega_\perp/(2 \pi)$ at a value close to 2.85~kHz. Using evaporative cooling, we obtain a cold sample at thermal equilibrium in the trap. Temperatures as low as 1.4~$\hbar \omega_\perp / \kB$ are accessible with an atom number of $5 \times 10^3$. The atomic cloud has a typical length of 100~$\mu$m along the $z$ axis and a transverse radius of 300~nm.

As shown in Fig.~\ref{fig.dispositif}, \emph{in situ} absorption
images are taken using a probe beam perpendicular to the $z$ axis
and reflecting on the chip surface at $45^{\rm o}$. The light,
resonant with the closed transition $|F=2\rangle \rightarrow
|F'=3\rangle$ of the D2 line is switched on for 150~$\mu$s with an
intensity of one tenth of the saturation intensity. Two images are
recorded with a CCD camera whose pixel size $\Delta \times \Delta$
in the object plane is $6.0 \times 6.0$~$\mu$m$^2$. The first image
is taken while the trapping field is still on. The second image is
used for normalization and is taken in the absence of atoms 200~ms
later. During the first image, the cloud expands radially to about
5~$\mu$m because of the heating due to photon scattering by the
atoms. The size of the cloud's image is even larger due to
resolution of the optical system (about 10~$\mu$m) and because the
cloud and its image in the mirror at the atom chip surface are not
resolved. Five pixels along the transverse direction $x$ are needed
to include 95\% of the signal.

We denote by $N^{\rm ph}_i(x,z)$ the number of photons detected in
the pixel at position $(x,z)$ for the image $i$ $(i=1,2)$. We need
to convert this measurement into an atom number $N(z)$ detected
between $z$ and $z+\Delta$. Normally, one simply computes an
absorption per pixel $\ln(N^{\rm ph}_2/N^{\rm ph}_1)$ and sums over
$x$:
\begin{equation}\label{eq.measured_density}
N(z) = \sum_x \, \ln[N^{\rm ph}_2(x,z)/N^{\rm ph}_1(x,z)] \times
\Delta^2/\sigma_e,
\end{equation}
where $\sigma_e$ is the absorption cross-section of a single atom.
When the sample is optically thick and the atomic density varies on
a scale smaller than the optical resolution or the pixel size,
Eq.~(\ref{eq.measured_density}) does not hold since the logarithm
cannot be linearized. In that case, Eq.~(\ref{eq.measured_density})
underestimates the atom number and the error increases with optical
thickness. 
 Furthermore, in our geometry, optical rays cross the atomic cloud
 twice 
since the cloud image and its reflexion in the atom chip surface 
are not resolved.

We partially correct for these effects by using in
Eq.~(\ref{eq.measured_density}) an effective cross section $\sigma_e$
determined as follows. We compare the measured atom number using the
\emph{in situ} procedure described above with the measured atom
number after allowing the cloud to expand
and to leave the vicinity of
the chip surface. In this case,
Eq.~(\ref{eq.measured_density}) is valid and the atomic
cross-section $\sigma_0=3\lambda^2/(2\pi)$ well known. We then
obtain for the effective cross-section $\sigma_e=0.8 \, \sigma_0$.
Although this effective cross section depends on the atomic density,
we have checked that for total atom number between $2\times 10^3$
and $9\times 10^3$ the measured value varies by only 10\%. Taking
into account the uncertainty on the value of $\sigma_0$, we estimate
the total error on the measured atom number $N(z)$ to be less than
20\%. 

\begin{figure}
\includegraphics{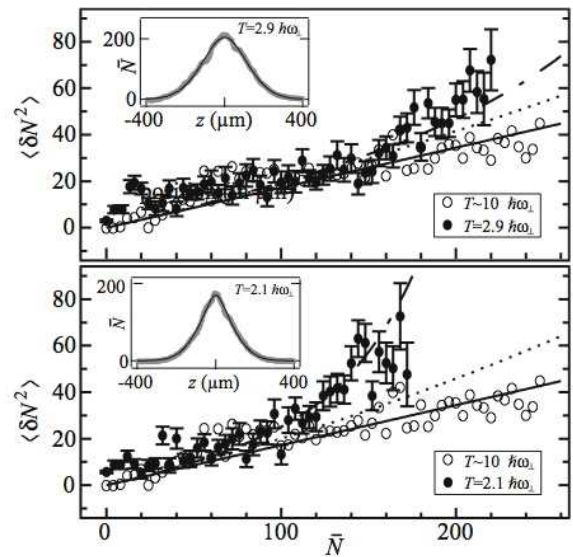}
\caption{Atom number variance as a function of the mean atom number
per pixel. Open circles correspond to a "hot" cloud ($\kB T\simeq 10 \,
\hbar \omega_\perp$, $\omega_\perp=2 \pi \times 2.85$~kHz) for which
fluctuations are given by the shot noise (black line). The full
circles correspond to cold clouds.
Error bars show the 
standard deviation of the mean of $\langle \delta N^2\rangle$.
 The fluctuations in excess of
shot noise are due to bosonic bunching. The dot-dashed line is the
prediction for an ideal Bose gas while the dotted line uses the
Maxwell-Boltzmann approximation (see Eq.~(\ref{eq.pixelfluctu})).
The insets show the longitudinal profile of the two cold clouds from
which we deduce the temperature and the chemical potential used for
the calculations.} \label{fig.bunching}
\end{figure}

\begin{figure}[t]
\includegraphics{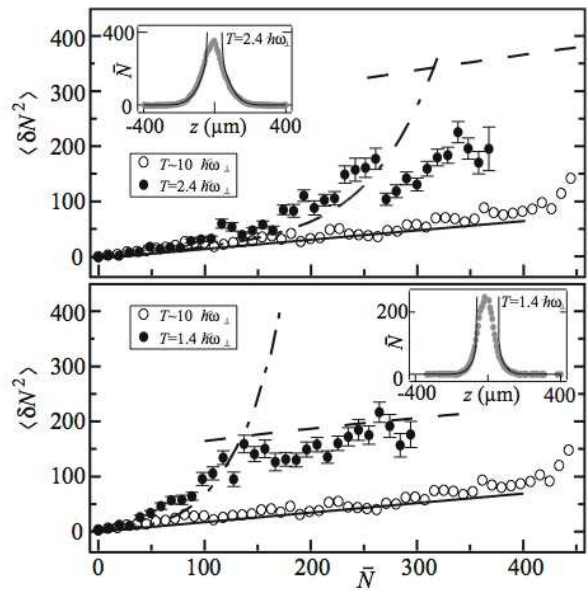}
\caption{We plot the same quantities as in
figure~\ref{fig.bunching}. Dot-dashed lines are the predictions for
an ideal Bose gas (deduced from Eq.~(\ref{eq.flucturhosansint})), 
whereas the dashed lines show the results of 
 Eq.~(\ref{eq.fluctuthermo}). The temperature of the sample is
deduced by fitting the wings of the longitudinal profile to an ideal
Bose gas profile as shown in the insets. 
The solid lines gives the atomic shot noise level.} \label{fig.fluctuquasibec}
\end{figure}

To measure the variance of the atom number in a pixel, we acquire a
large number of images (typically 300) taken in the same
experimental conditions. To remove technical noise from our
measurement, the following procedure is used to extract the
variance. For each image, we form the quantity $\delta N(z)^2 =(N
(z) - \bar{N}(z))^2$ where the mean value
$\bar{N}(z)$ is normalized to contain the same total atom number as
the current image. We thus correct for shot to shot total atom
number fluctuations. The average is performed only over $p=21$
images which bracket the current image so that long term drifts of
the experiment do not contribute to the variance. We have checked
that the results are independent of $p$, varying $p$ between 5 and
21~\footnote{We actually form the quantity 
$\delta N(z)^2 =(N(z) - \bar{N}(z))^2\times p/(p-1)$
to take into account the underestimation of the variance due to 
finite number of images.}.
A large contribution to $\delta N(z)^2$, irrelevant to our
study, is the photon shot noise of the absorption measurement. To
precisely correct for this noise, we subtract the quantity $\sum_x
(1/N^{\rm ph}_{1}(x,z)+1/N^{\rm ph}_{2}(x,z)) (\Delta^2\sigma_e)^2$
from $\delta N(z)^2$ for each image. We typically detect $10^4$
photons per pixel corresponding to a contribution to $\delta N^2$ of
about 50. To convert the camera signal into a detected photon
number, we use a gain for each pixel that we determine by measuring
the photon shot noise of images without atoms as explained
in~\cite{jiang:521}. The corrected $\delta N(z)^2$ obtained for all
images are then binned according to the value of $\bar{N}(z)$,
rather than of $z$ itself. This gives the variance of the atom
number $\langle \delta N(z)^2 \rangle$ as a function of the mean
atom number per pixel. Since more pixels have a small atom number,
the statistical uncertainty on the estimate of the variance decreases with
the average atom number (see Figs.~(\ref{fig.bunching})
and~(\ref{fig.fluctuquasibec})).

Data shown in Fig.~\ref{fig.bunching} correspond to atom clouds of
sufficiently low density so that effect of inter-atomic interactions is
expected to be small. The three data sets
correspond to three different temperatures, the trapping frequencies 
are 2.85~kHz and 7.5~Hz. We deduce the temperature and the chemical
potential of the sample by fitting the mean longitudinal profile
$\bar{N}(z)$ of the cloud to the profile of an ideal Bose gas (see
inset of Fig.~\ref{fig.bunching}).  For the "hot" sample where
bunching  gives negligible contribution to $\delta N^2$  
(see Eq.~(\ref{eq.pixelfluctu})), we observe
atomic shot noise fluctuations, \emph{i.e.} the atom number variance
increases linearly with the mean atom number. 
The fact that we recover the linear behavior expected for shot noise
increases our confidence in the procedure described in the previous
two paragraphs.
The slope $\kappa$ is
only 0.17 and differs from the expected value of 1. We attribute
this reduction to the fact that our pixel size is not much bigger
than the resolution of our optical imaging system, thus one atom is
spread out on more than one pixel. When the pixel size is small
enough compared to the optical resolution and in the case of weak
optical thickness, the expected slope is simply approximated by
$\kappa \simeq \Delta /(2 \sqrt{\pi} \, \delta)$ where $\delta$ is
the rms width of the optical response which we suppose gaussian. From
the measured slope, we deduce $\delta=10$~$\mu$m  in good
agreement with the smallest cloud image we have observed (8~$\mu$m).

For "cold" samples, we see an excess in the atom number variance
compared to shot noise. We attribute this excess to bunching due to
the bosonic nature of the atoms. In a local density approximation,
the fluctuations of a radially trapped Bose gas with longitudinal density $n(z)$  are~\cite{Glauber} 
\begin{equation}
\begin{array}[t]{l}
\langle n(z) \, n(z')\rangle -\langle n(z) \rangle^2 = \langle n(z)
\rangle \, \delta(z-z')  + \\ \displaystyle \frac{1}{\ldb^2}
\sum_{i=1}^\infty\sum_{j=1}^\infty  \frac{e^{\beta  \mu  (i+j) }}{\sqrt{i  j}} \frac{
e^{-\pi  (z-z')^2 (\frac{1}{i}+\frac{1}{j})/ \ldb^2 }} { \left [
1-e^{-\beta  \hbar  \omega_{\perp} (i+j)} \right ]^2 }
\end{array}
\label{eq.flucturhosansint}
\end{equation}
where $\mu$ is the local chemical potential, $\beta = 1/ (\kB T)$,
$\ldb =\sqrt{2 \pi \hbar^2/ (m  \kB  T)}$ is the de Broglie thermal
wavelength and $\langle . \rangle$ denotes an ensemble average. The
first term on the right hand side corresponds to shot noise, and the
second term to bunching. For a non degenerate gas $(n  \ldb \ll 1)$,
one can keep only the term $i=j=1$. The bunching term reduces to
$\langle n(z) \rangle^2 \exp(-2 \pi  (z-z')^2/\ldb^2) \, \tanh^2 (
\beta  \hbar  \omega_\perp/2 )$ and one recovers the well-known
gaussian decay of the correlations. 
The reduction factor $\tanh^2 (
\beta  \hbar  \omega_\perp/2 )$ is due to the integration over the
transverse states. In our experiment, the pixel size is always much
bigger than the correlation length. In which case, integrating over
the pixel size~$\Delta$, we have
\begin{equation}\label{eq.pixelfluctu}
\langle N^2 \rangle - \langle N \rangle^2 = \langle N \rangle +
\langle N \rangle^2 \frac{\ldb}{\sqrt{2}  \Delta} \tanh^2 ( \beta
\hbar  \omega_\perp/2 ).
\end{equation}
The coefficient of  $\langle N \rangle^2$ is the inverse of the
number of elementary phase space cells occupied by the $N$ atoms.

To compare Eq.~(\ref{eq.pixelfluctu}) to our data we must correct
for the optical resolution as was done for the shot noise.
Furthermore, atoms diffuse about 5~$\mu$m during the imaging pulse
because of photon scattering. This diffusion modifies the
correlation function, but since the diffusion distance is smaller
than the resolution, 10~$\mu$m, and since its effect is averaged
over the duration of the pulse, its contribution is negligible. We
thus simply multiply the computed atom number variance by the factor
$\kappa$.

Figure~\ref{fig.bunching} shows that the value calculated from
Eq.~(\ref{eq.pixelfluctu}) (dotted line) underestimates the observed
atom number variance. In fact, for the coldest sample, we estimate
$n(0) \ldb \simeq 10$, and thus the gas is highly degenerate. In
this situation replacing the Bose-Einstein occupation numbers by
their Maxwell-Boltzmann approximations is not valid, meaning that
many terms of the sum in Eq.~(\ref{eq.flucturhosansint}) have to be
taken into account. The prediction from the entire sum is shown as a
dot-dashed line and is in better agreement with the data.

In the experiment we are also able to access the quasi condensate regime in which inter-particle interactions are not negligible, and the ideal gas
theory discussed above fails. Figure~\ref{fig.fluctuquasibec} shows
the results of two experimental runs using denser clouds.
 For these data, the trapping frequencies are 2.85~kHz and
10.5~Hz. The insets show the mean longitudinal cloud profiles and a
fit to the wings of the profiles to an ideal Bose gas profile. One
can see from these insets that, unlike the conditions of
Fig.~\ref{fig.bunching}, an ideal gas model does not describe the
density profile in the center. We employ the same procedure to
determine the variance versus the mean atom number. As in
Fig.~\ref{fig.bunching} we plot our experimental results along with
the ideal Bose gas prediction based on the temperature determined
from the fit to the wings in the insets. For small mean value
$\bar{N}(z)$, the measured fluctuations follow the ideal gas curve
(dot-dashed line) but they are dramatically reduced when the atom
number is large.

The theory for a weakly interacting uniform 1D Bose gas permits an
analytical prediction for the density fluctuations in the limit
$n\gg n_T$. In this limit, the gas enters the Gross-Pitaevskii
regime and density fluctuations are given in the Bogoliubov
approximation by~\cite{Kheru2003,Quasibec_Castin}
\begin{equation}
\begin{array}[t]{l}
\langle \delta n(z) \, \delta n(z')\rangle = \\
\displaystyle \frac{\langle n\rangle}{2 \pi}
\int_{-\infty}^{\infty} dk\, e^{ik(z-z')} \left(\frac{k^2}{k^2+ 4
\xi^{-2} }\right)^{1/2} (1+2 n_k), \label{eq.deltarhoquasibec}
\end{array}
\end{equation} 
where $n_k$  is the Bose thermal occupation factor of the mode $k$
with energy $\epsilon_k=\sqrt{k^2 (k^2 + 4 \xi^{-2})} \times
\hbar^2/(2m)$ and $\xi=\hbar/\sqrt{m  n  g}$ is the healing length.
For 200 atoms per pixel, the healing length is about 0.3~$\mu$m 
in our experiment~\footnote{The phase correlation
length $l_c=\hbar^2 n/(m \kB T)$ is about 1~$\mu$m for $\kB
T=1.4\,\hbar\omega_{\perp}$ confirming that we are in the quasi-condensate regime.}. The term proportional to $n_k$ describes the contribution of thermal
fluctuations while the other is due to vacuum fluctuations. Since
the pixel size is much bigger than the healing length, we probe only
long wavelength fluctuations for which thermal fluctuations dominate
at the temperatures we consider. Using $k \ll 1/\xi$ and $n_k \simeq
\kB  T / \epsilon_k$, we obtain for the atom number variance in a
pixel
\begin{equation}\label{eq.QuasiBEC}
\langle N^2\rangle-\langle N\rangle^ 2=\Delta \frac{\kB  T}{g}.
\end{equation}
This formula can also be deduced from thermodynamic considerations:
for a gas at thermal equilibrium, the atom number variance in a
given volume is given by
\begin{equation}\label{eq.thermo}
\langle N^2\rangle-\langle N\rangle^ 2=\kB T (\partial N / \partial
\mu)_T.
\label{eq.fluctuthermo}
\end{equation}
For a quasi-condensate with chemical potential $g n$,
equations~(\ref{eq.QuasiBEC}) and~(\ref{eq.thermo}) coincide.

The calculation leading to Eq.~(\ref{eq.QuasiBEC}) holds in a true
1D situation in which case the effective coupling constant is
$g=2\hbar \omega_\perp a$, where $a$ is the scattering length of the
atomic interaction. The validity condition for the 1D calculation is
$n \ll 1/a$ (equivalently $\mu \ll \hbar \omega_\perp$). In our
experiment however, the value of $n a$ is as high as 0.7 and thus
one cannot neglect dependence of the transverse profile on the local
density. On the other hand, the thermodynamic approach 
 is valid and, supposing $\mu(N)$ is
known,
Eq.~(\ref{eq.thermo}) permits a very simple calculation. We use the approximate
formula $\mu(N) = \hbar \omega_\perp \sqrt{1 + 4 N a/\Delta}$ valid
in the quasi-condensate regime~\cite{gerbier:771}. This formula
connects the purely 1D regime with that in which the transverse
profile is Thomas-Fermi. The results of this analysis, confirmed by
a full 3D Bogoliubov calculation, are plotted in
Fig.~\ref{fig.fluctuquasibec} (dashed line).
Equation~(\ref{eq.QuasiBEC}) predicts a constant value for the atom
number variance and underestimate it by 50\% for the maximal density
reached in our experiment ($\bar{N}=400$).

We compare this calculation in the quasi-condensate regime with our
data. From Fig.~\ref{fig.fluctuquasibec} we see that the calculation
agrees well with the measurements for $\kB T=1.4 \, \hbar \omega_\perp$
but less so for $\kB T=2.4 \, \hbar \omega_\perp$.  The one-dimensional theory predicts that  the quasi-condensate approximation is valid in the limit $n\gg n_T$ which corresponds to $\langle N \rangle \gg $ 100 (140) for $\kB T=1.4\,
\hbar \omega_\perp$ ($\kB T=2.4 \, \hbar \omega_\perp$).
The disagreement between the calculation and our data for $\kB T=2.4 \,
\hbar \omega_\perp$
suggests that perhaps we did not achieve a high enough density 
to be fully in the quasi-condensate approximation.
In addition the one-dimensional calculation of $n_T$ is unreliable 
for such high ratio $\kB T/\hbar\omega_\perp$ and underestimates the
value at which the cross over appears. This is also the case for the 
data of Fig.2 where the naive estimate of $n_T$ corresponds to 
$\langle N \rangle =160$ for $\kB T=2.9\hbar\omega_\perp$
and $\langle N \rangle=130$ for $\kB T=2.1\hbar\omega_\perp$.

Exploitation of the  1D geometry to avoid averaging the fluctuations in the imaging direction  can be applied to other situations. A  Bose gas in the strong coupling regime, or an elongated Fermi gas should show sub shot noise fluctuations due to anti-bunching.

This work has been supported by the EU under grants IST-2001-38863, MRTN-CT-2003-505032, IP-CT-015714, by the DGA (03.34033) and by the French research ministry  ``AC nano'' program. We thank D.~Mailly from the LPN (Marcoussis,
France) for helping us to micro-fabricate the chip.

\end{document}